\documentclass[aps,prb,twocolumn,groupaddress]{revtex4-1}
\usepackage{amsmath}
\usepackage{amssymb}
\usepackage{mathtools}
\usepackage{graphicx}
\usepackage{hyperref}
\graphicspath{ {./figs/} }
\usepackage{tabularx}
\usepackage{bm}
\usepackage{amsthm}
\usepackage[toc,page]{appendix}
\usepackage{enumitem}
\usepackage{xcolor}

\newcommand{\Hsh}{$H_{\textrm{sh}}$}

\begin{document}

\title{Role of surface defects and material inhomogeneities for vortex nucleation in superconductors within time-dependent Ginzburg-Landau theory in 2 and 3 dimensions}


\author{Alden R. Pack}
\email{a.pack@byu.edu}
\author{Jared Carlson}
\author{Spencer Wadsworth}
\author{Mark K. Transtrum}
\email{mktranstrum@byu.edu}
\affiliation{Department of Physics and Astronomy, Brigham Young University, Provo, Utah 84602, USA}

\date{\today}

\begin{abstract}
  We use Time-Dependent Ginzburg-Landau theory to study the nucleation of vortices in type II superconductors in the presence of both geometric and material inhomogeneities.
  The superconducting Meissner state is meta-stable up to a critical magnetic field, known as the superheating field.
  For a uniform surface and homogenous material, the superheating transition is driven by a non-local critical mode in which an array of vortices simultaneously penetrate the surface.
  In contrast, we show that even a small amount of disorder localizes the critical mode and can have a significant reduction in the effective superheating field for a particular sample.
  Vortices can be nucleated by either surface roughness or local variations in material parameters, such as $T_c$.
  Our approach uses a finite element method to simulate a cylindrical geometry in 2 dimensions and a film geometry in 2 and 3 dimensions.
  We combine saddle node bifurcation analysis along with a novel fitting procedure to evaluate the superheating field and identify the unstable mode.
  We demonstrate agreement with previous results for homogenous geometries and surface roughness and extend the analysis to include variations in material properties.
  Finally, we show that in three dimensions, suface divots not aligned with the applied field can increase the super heating field.
  We discuss implications for fabrication and performance of superconducting resonant frequency cavities in particle accelerators.
\end{abstract}

\maketitle

\section{Introduction}


A hallmark feature of type-II superconductors is a phase transition from a purely superconducting (i.e., Meissner) state to a mixed state characterized by arrays of magnetic vortices.
The mixed state can be understood as the compromise in the competition between magnetic pressure and the condensation of Cooper pairs.
If the characteristic length scales for these phenomena are appropriately separated, a balance is struck in which filaments of magnetic field and small, non-superconducting cores are trapped by vortices of supercurrent.
This configuration is thermodynamically stable between a lower and upper critical field ($H_{c1}$ and $H_{c2}$ respectively).
Olsen et. al. beautifully captured this behavior using magneto-optical imaging \cite{olsen2004single}.
For time-independent configurations, a stable array of vortices can be achieved, while for alternating magnetic fields, vortex motion leads to heat dissipation\cite{liarte2018vortex}.

Ginzburg-Landau (GL) theory succinctly captures the relevant physics for describing the Meissner and vortex states, as well as the transition between the two.
The theory is described by two characteristic length scales, the London penetration depth $\lambda$ and the superconducting coherence length $\xi$.
For materials in which the ratio $\kappa=\lambda/\xi$ (known as the GL parameter) is less than $1/\sqrt{2}$ the material is type I and will transition directly from the Meissner state to the nonsuperconducting state.
However, for type II superconductors ($\kappa>1/\sqrt{2}$) the material transitions first to a mixed, vortex state.
The density of vortices increases with larger applied magnetic field until the system transitions to a nonsuperconducting state at $H_{c2}$.

Although vortices are thermodynamically stable for fields above $H_{c1}$, surface effects lead to an energy barrier to vortex nucleation\cite{bean1964surface}.
The Meissner state can persist above $H_{c1}$ up to a maximum magnetic field, known as the \emph{superheating field} $H_{sh}$  above which the energy barrier vanishes.
For homogenous materials with smooth surfaces, this transition is driven by critical perturbations with a characteristic wavenumber $k_c$.
For applications requiring a Meissner state (i.e., for which vortex nucleation is detrimental), $H_{sh}$ is the fundamental limit to performance.
As such, estimates of $H_{sh}$ within Ginzburg-Landau theory have a long history \cite{transtrum2011superheating, chapman1995superheating, dolgert1996superheating,kramer1968stability,de1965vortex,galaiko1966stability,kramer1973breakdown,fink1969stability,christiansen1969magnetic}.
This technique has since been extended to Eilenberger theory in both the clean\cite{catelani2008temperature} and dirty\cite{lin2012effect} limits.
Often real systems have rough surfaces and interior defects that don't match this geometry.
The role of surface roughness on $H_{sh}$ in two dimensional geometries with surface defects has been studied extensively within Ginzburg-Landau theory \cite{soininen1994stability, vodolazov2000effect, burlachkov1991bean, aladyshkin2001best}.
There has also been considerable effort to simulate vortex nucleation and subsequent dynamics for more complicated domains within time-dependent Ginzburg-Landau (TDGL) theory\cite{machida1993direct, koshelev2016optimization, du1994finite, oripov2019time, dorsey1992vortex, li2015new, gao2015efficient, sadovskyy2015stable,sorensen2017dynamics,deang1997vortices,benfenati2019vortex}.

Particle accelerators are an application of importance to a wide variety of fields\cite{hamm2012industrial, amaldi1999cancer, malumud2010accelerators} to which quantitative studies of the superheating field and vortex motion are particularly relevant.
Superconducting Radio Frequency (SRF) cavities transfer energy to particle beams.
Large AC currents running along the interior surface of the cavity induce electromagnetic fields that are timed to boost particle bunches as they pass through \cite{padamsee2008rf}.
Traditionally cavities have been fabricated from Nb, but engineering advances have pushed these cavities to near their fundamental limits\cite{posen2015radio}.

To more efficiently reach higher accelerating gradients, the accelerator community is exploring new materials for next-generation cavities\cite{posen2013srf}.
Of particular interest is Nb$_3$Sn, which theoretically has $H_{sh}=425[mT]$ and $T_c=18[K]$ (compared to Niobium which has $H_{sh}=219[mT]$ and $T_c=9.2[K])$\cite{posen2015understanding}.
In practice current Nb$_3$Sn cavities perform far-below theoretical limits\cite{posen2015radio, posen2014advances}.

In addition to surface roughness, the alloyed nature of these materials often leads to variations in material parameters, such as Sn concentration, that can have a strong effect on the superconducting properties\cite{devantay1981physical,devantay1982superconductivity,godeke2006review,becker2015analysis, lee2018atomic, sitaraman2019textit}.
To guide future development and keep pace with experimental advancements, more sophisticated theoretical and computational tools are needed to identify the relevant physics for vortex nucleation and quantify their effect on $H_{sh}$ in real materials.
They need to be flexible enough to not only capture the impact of surface roughness, but also interior material inhomogeneities.
These advances also offer an opportunity to validate theories of traditional superconductors in extreme conditions.

In this paper, we perform bifurcation analysis of the Meissner state using TDGL and a finite-element formulation.
Our method quantitatively confirms previous estimates of $H_{sh}$ derived in the symmetric, time-independent theory.
We account for asymmetric geometries, such as surface divots, and variation in material parameters in two and three dimensions.
We show a that local reductions in the superconducting critical temperature is a potentially important nucleation mechanism in inhomogenous alloyed superconductors.
Our method identifies the critical fluctuations that drive the vortex nucleation.
Unlike the symmetric case in which arrays of vortices nucleate in tandem, a small amount of disorder acts as a nucleation site for individual vortices, indicating that near $H_{sh}$, the free energy surface has several shallow directions.
We quantify this effect for both surface roughness and material inhomogeneity, a result that will guide the manufacture of precision samples to maximize performance.
Finally, in three dimensions we show that the relative orientation of defects and the external field has a strong role in vortex nucleation.
We demonstrate that defects aligned perpendicular to the applied field lead to an increase in $H_{sh}$.


The rest of this paper is organized as follows.
Section \ref{Methods} formulates the time-dependent Ginzburg-Landau (TDGL) equations to account for spatial variations in $T_c$ and introduces the two- and three-dimensional geometries we consider.
We also introduce saddle-node bifurcation analysis to efficiently identify the critical modes that drive vortex nucleation and estimate $H_{sh}$.
In section \ref{Results} we first confirm that our simulations for homogenous systems match previous work.
Then we report on the effect of surface roughness and material inhomogeneity in two and three dimensions.
Finally, in section~\ref{sec:Summary}, we discuss implication and limitations of our approach and potential future extensions.

\section{Methods}\label{Methods}

\subsection{Problem Formulation}\label{Problem Formulation}
The time-dependent Ginzburg-Landau (TDGL) equations are a series of partial differential equations relating the superconducting order parameter to the electric potential and magnetic vector potential on mesoscopic scales.
Although originally a phenomenological theory, the equations can be rigorously derived from the time-dependent Gorkov equations\cite{gor1996generalization}.
The TDGL equations in Gaussian units given in ref.\cite{kopnin2001theory} are
\begin{widetext}
\begin{align}
  -\Gamma\Big(\frac{\partial\psi}{\partial t} + \frac{2ie\phi}{\hbar}\psi\Big) = & -|\alpha|\psi + \beta|\psi|^2\psi + \gamma\Big(-i\hbar\nabla-\frac{2e}{c}\mathbf{A}\Big)^2\psi\\
  \mathbf{j} = & \frac{c}{4\pi}\nabla\times\nabla\times\mathbf{A}\nonumber \\
                                                                                    =& \sigma_n\Big(-\frac{1}{c}\frac{\partial\mathbf{A}}{\partial t} - \nabla\phi\Big)+2e\gamma\Big[\psi^*\Big(-i\hbar\nabla-\frac{2e}{c}\mathbf{A}\Big)\psi+\psi\Big(i\hbar\nabla-\frac{2e}{c}\mathbf{A}\Big)\psi^*\Big].
\end{align}
\end{widetext}
These equations depend on the order parameter $\psi$, the magnetic vector potential $\mathbf{A}$, and the electric potential $\phi$ all of which can vary in space and time.
The rest of the quantities are materials parameters and fundamental constants: $\Gamma$ is the rate of relaxation of the order parameter, $e$ is the charge of an electron, $\hbar$ is Plank's constant divided by 2$\pi$, $c$ is the speed of light, $\alpha$ is a material-specific constant proportional to $1-T/T_c$ ($T$ is temperature and $T_c$ is the critical temperature), $\beta$ is anther material parameter that is approximately constant with respect to $T_c$, $\gamma$ is related to the effective mass of the cooper pairs, and $\sigma_n$ is the conductivity of the normal electrons.

Typically, all physical constants can be absorbed into the units of fields.
However, we relax this assumption in order to model spatial variations in $T_c$ by allowing $\alpha(r) \propto 1 - T/T_c$ to vary in space over a range of values.
This has been done previously to model pinning sites by setting $\alpha(r)$ to zero at fixed points in the domain\cite{koshelev2016optimization,sadovskyy2015stable,sorensen2017dynamics,deang1997vortices}.
We define $\alpha(r) = \alpha_0 a(r)$ where $\alpha_0$ is a reference value (to be subsumed by units), and $a(r)$  is a dimensionless number characterizing the spatial material variation.
The quantities $\alpha_0$ and $a(r)$ are defined with respect to some reference point in the bulk material such that $a(r_0)=1$ and $\alpha(r_0) = \alpha_0$.
With this convention, $\alpha_0$ can be absorbed into the units of the field.
Values of $a$ less than one correspond to a local $T_c$ less than the reference value with $a < 0$ corresponding to $T_c$ less than the operating temperature.
The critical temperature of Nb$_3$Sn can depend strongly on the local concentration of Sn\cite{sitaraman2019textit}, so local reductions in $T_c$ are an important potential mechanism for vortex nucleation.

With these modifications and assuming our boundary conditions are a fixed applied magnetic field on the surface with no current leaking into vacuum, we arrive at
\begin{widetext}
\begin{align}
  \frac{\partial \psi}{\partial t} + i \phi\psi = & -a\psi + |\psi|^2\psi+\left(\frac{-i}{\kappa_0}\nabla- \mathbf{A}\right)^2\psi\label{eq:TDGL1}\\
  \mathbf{j}= & \nabla\times{\nabla\times{\mathbf{A}}}\nonumber\\
                                                  = &- \frac{1}{u_0}\left(\frac{\partial \mathbf{A}}{\partial t} + \frac{1}{\kappa_0}\nabla\phi\right)-\frac{i}{2\kappa_0}\left( \psi^* \nabla\psi - \psi\nabla\psi^*\right) - |\psi|^2\mathbf{A}\label{eq:TDGL2}\\
    \left( \frac{i}{\kappa_0}\nabla\psi + \mathbf{A}\psi \right)\cdot n = & 0 \text{ on surface}\label{eq:TDGL3}\\
    \left(\nabla\times{\mathbf{A}}\right)\times n = & \mathbf{H}\times n \text{ on surface}\label{eq:TDGL4}\\
    -\left( \nabla\phi + \frac{\partial \mathbf{A}}{\partial t} \right)\cdot n = & 0 \text{ on surface}\label{eq:TDGL5},
\end{align}
\end{widetext}
where we have introduced two new constants, $u_0$ and $\kappa_0$.
The constant $u_0=\tau_\psi/\tau_j$ is the ratio of the timescales for variations in the order parameter and the current.
They are defined as,
\begin{align}
  \tau_\psi &= \frac{\Gamma}{|\alpha_0|}\\
  \tau_j &=\frac{\beta\sigma_n}{8e^2\gamma|\alpha_0|}=\frac{\sigma_n}{8e^2\gamma\psi_0^2}.
\end{align}
The constant $\kappa_0$ is the Ginzburg-Landau parameter, the ratio of the penetration depth $\lambda_0$ and the coherence length $\xi_0$.
All of these are defined with respect to the reference point $r_0$.

Eqs.\eqref{eq:TDGL1}-\eqref{eq:TDGL5} are a set of coupled partial differential equations in three dimensions.
A common simplification is to assume a symmetry in the z-direction and only consider variations in the x-y plane.
This assumption leads to a two-dimensional formulation which greatly reduces the computational overhead, but does limit the types of geometries that can be simulated.
We perform both two-dimensional and three-dimensional simulations in this paper.

We numerically solve the TDGL equations using a finite element method (FEM) implemented in FEniCS\cite{alnaes2015fenics}.
Because the TDGL equations are diffusion-like, the time-step is implemented through an implicit formula.
We use a backwards Euler formula, but higher order backwards difference formulas could also be applied.
A more detailed description of previous methods is given by Gao et. al.\cite{gao2015efficient}.

One reason for the large variety of FEM formulations is the need to choose a gauge.
Although physical quantities should remain the same in different gauges, the efficiency and accuracy of numerical methods with each gauge varies.
We follow the formulations and conventions of Gao et. al.\cite{gao2017efficient,gao2015efficient}.
Although the TDGL equations are nonlinear, by using solutions from the previous time steps, each time step can be formulated as a series of \emph{linear} equations.
For the two-dimensional case, the problem can be  reduced to a series of Laplace and diffusion equations of coupled scalar fields which we implement as Lagrange elements.
In three dimensions, the problem also reduces to a series of linear equations; however, the geometric nature of the magnetic field and vector potential in 3D require they be modeled as Ravier-Thomas and Nedelec elements of different orders.
The complexity of the three-dimensional formulation incurs a substantial computational cost (both in time and memory).

\begin{figure}
    \centerline{\includegraphics[width=0.9\columnwidth]{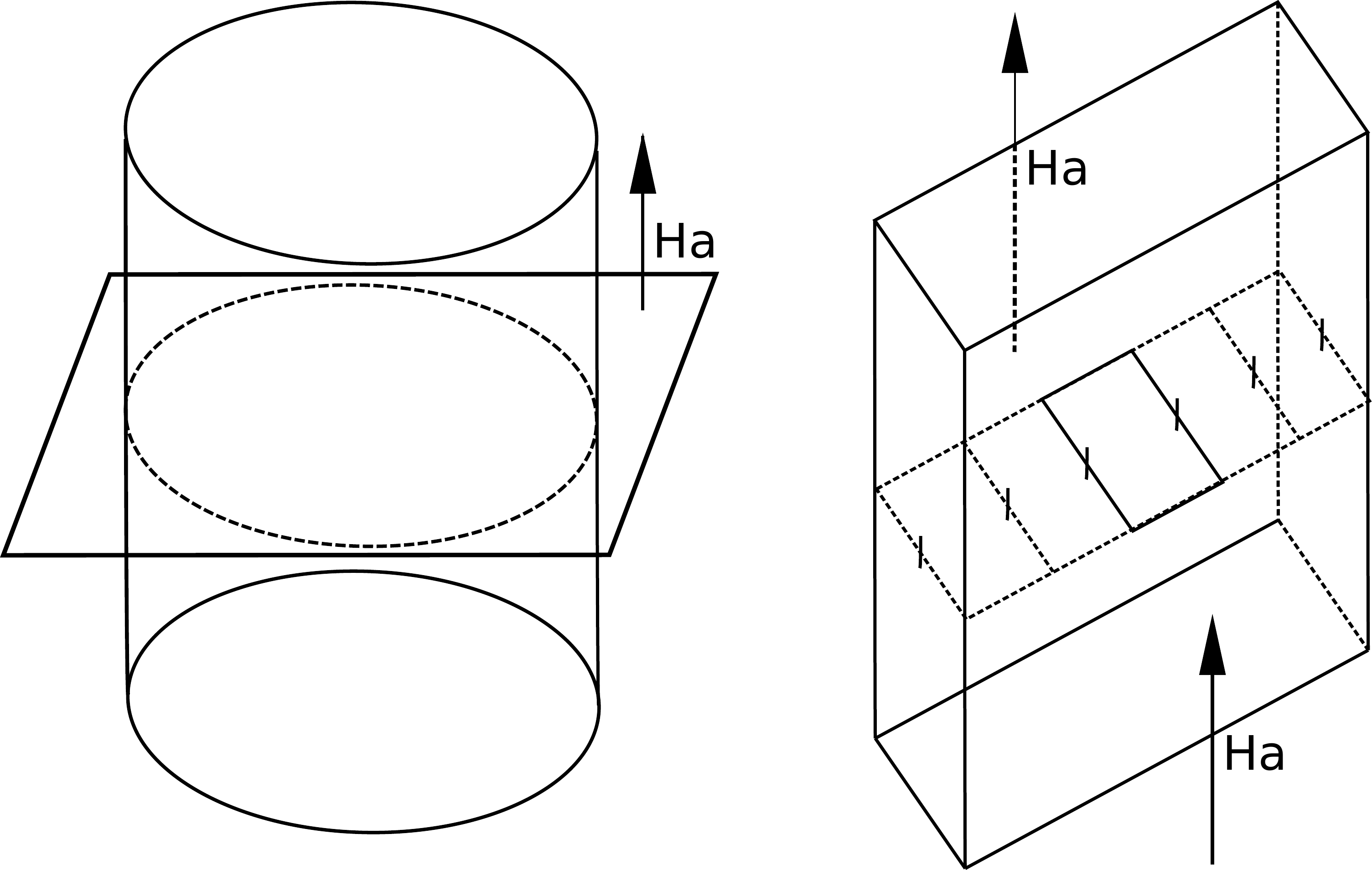}}
    \caption[CylandFilm]{\label{fig:CylandFilm}
      \textbf{Two Dimensional Geometries.}
      We consider an infinite superconducting cylinder (left) and an infinite superconducting film (right).
      In both cases, the magnetic field is perpendicular to the plane of simulation and does not vary spatially.
      Boundary conditions require matching the applied magnetic field on the surface.
      For the film (right), we have periodic boundary conditions on the left and right sides.}
\end{figure}

In the two dimensional case, we define two geometries: an infinite cylinder and a thin film.
In these geometries the magnetic field points in the $\hat{z}$ direction, i.e., perpendicular to the plane of simulation.
Fig.~\ref{fig:CylandFilm} show these cross sections.
For large radii and wide films these geometries approximate an infinite flat surface, studied using linear-stability analysis in reference\cite{transtrum2011superheating}.

\begin{figure}
    \centerline{\includegraphics[width=0.38\columnwidth]{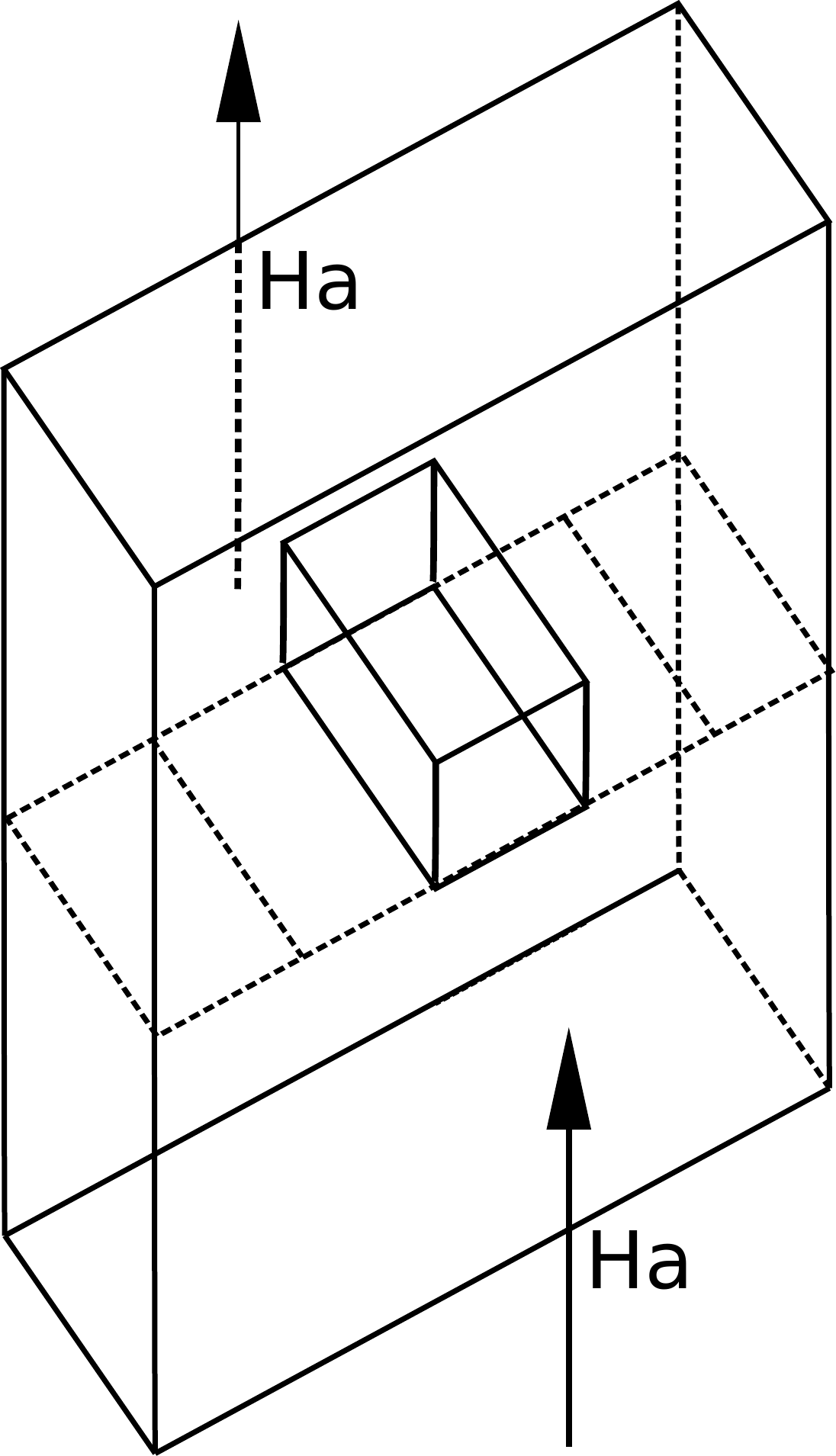}}
    \caption[Film3D]{\label{fig:Film3D}
      \textbf{Three Dimensional Geometry.}
      We generalize the 2D film geometry by extending the x-y plane along the $\hat{z}$ direction.
      In this geometry we are free to rotate the direction of the magnetic field.
    }
\end{figure}

In the 3D case we consider a rectangular box cut out of a thin film as in Fig.~\ref{fig:Film3D}.
This is done by extending the domain of simulation along the z axis (the inner solid box).
In this geometry we can orient the applied magnetic field in many directions along the surface of the film. 
The process of meshing these geometries is given in the appendix.

We take as initial conditions the case of a perfectly superconducting material in the absence of an applied field.
We raise the applied magnetic field exponentially to values near $H_{sh}$ in order to capture the dynamics of vortex nucleation.
The time dependence of the magnetic field is $\mathbf{H_a}(t)=\mathbf{H}_{max} (1-e^{-t/\tau})$.
This allows us to quickly raise the field but slow down close to the asymptotic value $\mathbf{H}$ where vortex nucleation is sensitive to small fluctuations in $\psi$ and $\mathbf{A}$.

\subsection{Inhomogeneities}\label{Inhomogeneities}

This formulation allows for a wide variety of potential simulations. 
We go beyond the bulk geometry \cite{transtrum2011superheating, chapman1995superheating, dolgert1996superheating,kramer1968stability,de1965vortex,galaiko1966stability,kramer1973breakdown,fink1969stability,christiansen1969magnetic} by considering the influence of surface roughness and spatial variations of $T_c$ ($\alpha$).

We introduce surface roughness in two ways.
First we model the surface of a wire (cylinder) as a Gaussian process (random sum of sinusoidal functions). 
Second, motivated by observed morphology of grain boundaries\cite{posen2017nb3sn}, we introduce a divot with a cutout of the form $Ae^{-|x| / \sigma}$.
Examples of these geometries are shown along with results in the next section and are described further in the appendix.

\begin{figure}
    \centerline{\includegraphics[width=1.0\columnwidth]{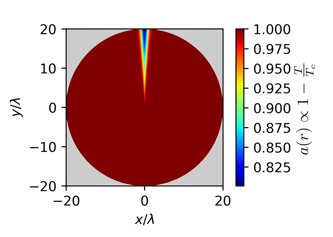}}
    \caption[Alpha Cyllinder]{\label{fig:Alpha Cylinder}
      \textbf{Spatial Dependence of a(r).}
      The dependence of the GL equations on the critical temperature comes from a coefficient $a$.
      We model the influence of Sn segregation as a local suppression of the superconducting critical temperature by allowing $a$ to vary spatially.
      Here we show the value of $a$ throughout the domain.
      $a<1$ leads to a reduction of the superconducting order parameter.
    }
\end{figure}

We model spatial variations of $T_c$ within the cylindrical geometry as a Gaussian function $a(r,\theta)=1-Be^{\frac{-\theta^2}{2s^2}}(\frac{r}{R})^l$, see Figure~\ref{fig:Alpha Cylinder}.
$B$ is the lowest value of alpha, $s$ sets the width of our defect, $R$ is the cylinder radius, and $l$ adjusts how quickly $a$ drops off radially.
This ``line'' of lowered $T_c$ mimics the effect of Sn segregation in the grain boundaries of Nb$_3$Sn cavities\cite{jewell2004upper}.

\subsection{Bifurcation Analysis and Mode Extraction}\label{Bifurcation}

One of the contributions of this work is a method for calculating \Hsh\ for arbitrary geometries and material properties.
The superheating field occurs when the meta-stable Meissner state becomes unstable to a critical fluctuation.
At \Hsh, the free energy landscape near the Meissner state transitions from a local minimum to a saddle point, and dynamics exhibit a saddle-node bifurcation.
The free energy flattens (to lowest order) in the direction characterizing the critical fluctuation that nucleates magnetic vortices.
Because the free energy landscape is flat near the bifurcation, simulation dynamics are slow for applied fields near \Hsh.
Rather than solve the TDGL equations near the bifurcation, we use normal-form theory to quickly extract \Hsh\ from simulations with applied fields below \Hsh.

The normal form of the saddle-node bifurcation is
\begin{equation}
  \label{eq:normalform}
 \frac{dx}{dt}=-r+x^2
\end{equation}
where $r$ is the bifurcation parameter\cite{strogatz2014nonlinear} and, in our case, an implicit, unknown function of the applied field.
$x$ is some combination of finite element degrees of freedom that becomes the unstable, critical fluctuation.

Eq.~\eqref{eq:normalform} is stable for $r>0$ and unstable for $r<0$.
Near the bifurcation, the system decays to equilibrium with a characteristic rate $\gamma=\frac{1}{2\sqrt{r}}$.
We extract the critical mode, $x$ by first finding the meta-stable Meissner state for applied fields below \Hsh.
We then perturb the state with random white noise and extract the slowest mode and the decay rate $\gamma$ using a fitting procedure\cite{pack2017computational}.
Repeating this calculation for several different applied fields, we then extrapolate to find the applied field at which $r$ becomes zero.
We also apply an iterative technique to improve the numerical stability of this method.
We repeatedly amplify the remaining noise and relax the system to cleanly separate the decaying mode and identify $\gamma$ and $r$ \cite{Carlson2019simulations}.
One of the benefits of this method is that it avoids running simulations where $r\approx0$ and the timescale diverges.

\section{Results}\label{Results}

\subsection{Agreement with Previous Work}

\begin{figure}
    \centerline{\includegraphics[width=1.0\columnwidth]{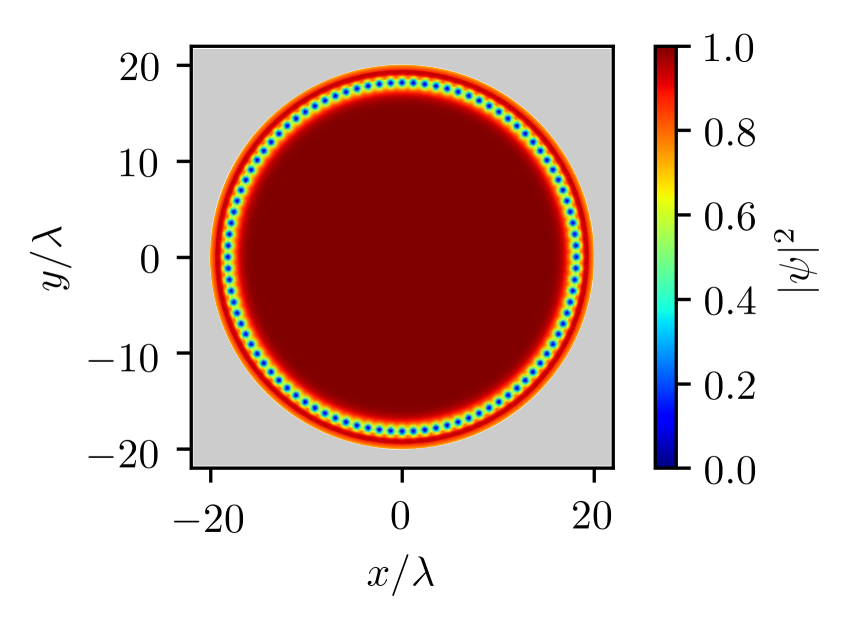}}
    \caption[Symmetric Vortices]{\textbf{Vortex Nucleation.}\label{fig:Symmetric Vortices}
        The order parameter above \Hsh\ after vortex nucleation.
        Note how the vortices penetrate uniformily around the cylinder.
    }
\end{figure}

We first demonstrate that our formulation correctly reproduces several known qualitative and quantitative results.
We reproduce vortex nucleation and numerical estimates of \Hsh\ using a cylindrical geometry without defects.
Fig.~\ref{fig:Symmetric Vortices} illustrates magnetic vortices shortly after nucleation for an applied magnetic field of $H_a=0.8\sqrt{2}H_c$ and a cylinder of radius $20\lambda$ with $\kappa=4$.  
Note that magnetic fields will always be measured in units of $\sqrt{2}H_c$ where $H_c$ is the thermodynamic critical field.
We will drop the $\sqrt{2}H_c$ from now on.

\begin{figure}
    \centerline{\includegraphics[width=1.0\columnwidth]{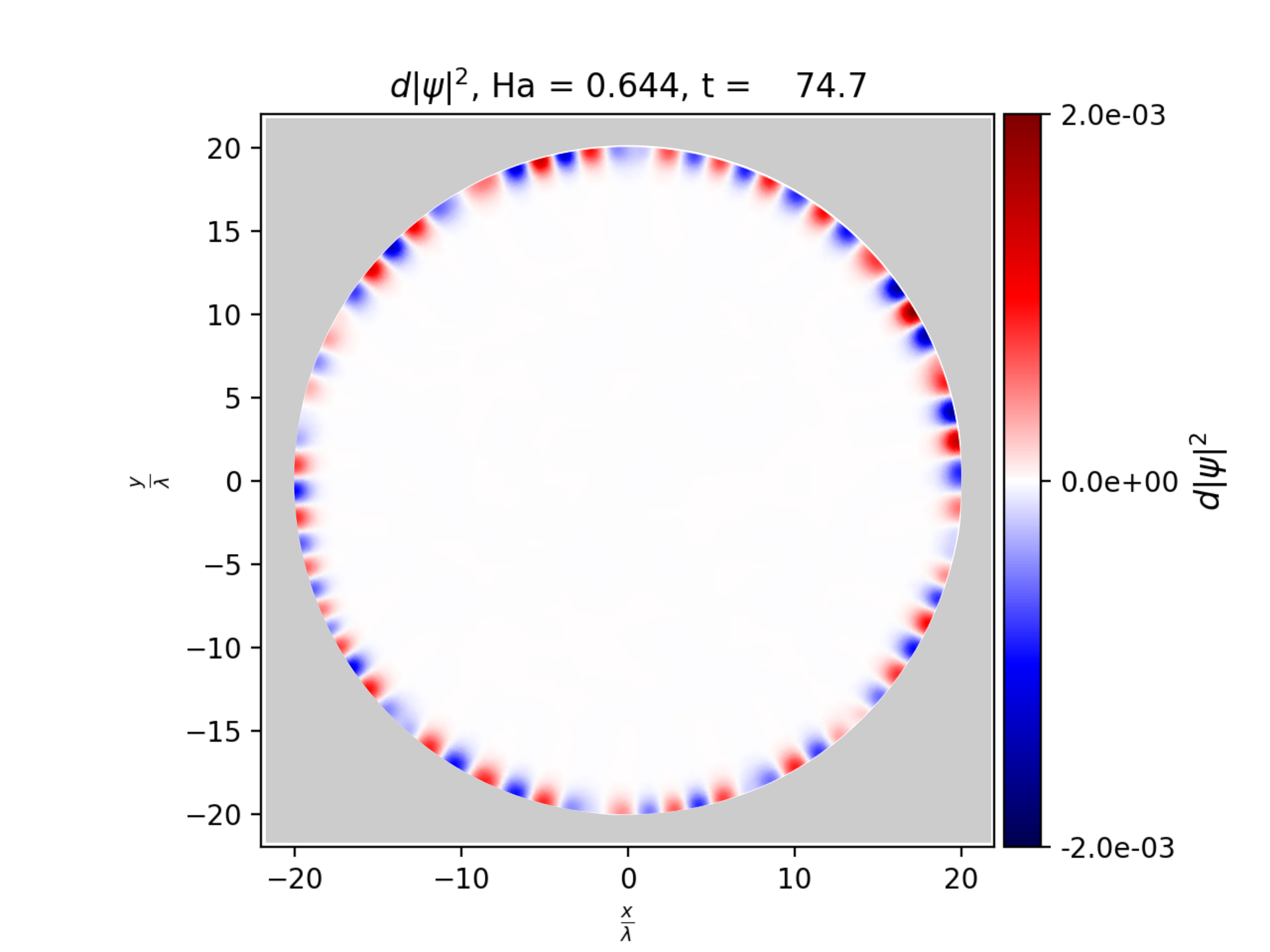}}
    \caption[Symmetric mode]{\label{fig:Symmetric mode}
      \textbf{Critical Fluctuation.}
      During the relaxation back to steady state after a random perturbation ($H_a < $\Hsh), the slowest decaying mode is the critical fluctuation that drives the phase transition at \Hsh.
      Note that the alternating pattern of low and high values roughly match the pattern of vortices in Figure~\ref{fig:Symmetric Vortices} and previous calculations of $k_c$ in bulk geometries.
    }
\end{figure}

As described in section \ref{Bifurcation} we extract the slowest decaying mode for fluctuations in the order parameter below but near $H_{sh}$.
Fig. \ref{fig:Symmetric mode} shows this mode for a radius of $20\lambda$.
This pattern is roughly sinusoidal on the surface wth a wavenumber $k_c$ that we estimate from the number of times the pattern crosses zero.
Fluctuations in this mode drive the transition from the Meissner state to the vortex state.
Notice that the mode is non-local.
The coordination of multiple penetrating magnetic votices lowers the barrier to entry for any single vortex.

\begin{figure}
    \centerline{\includegraphics[width=1.0\columnwidth]{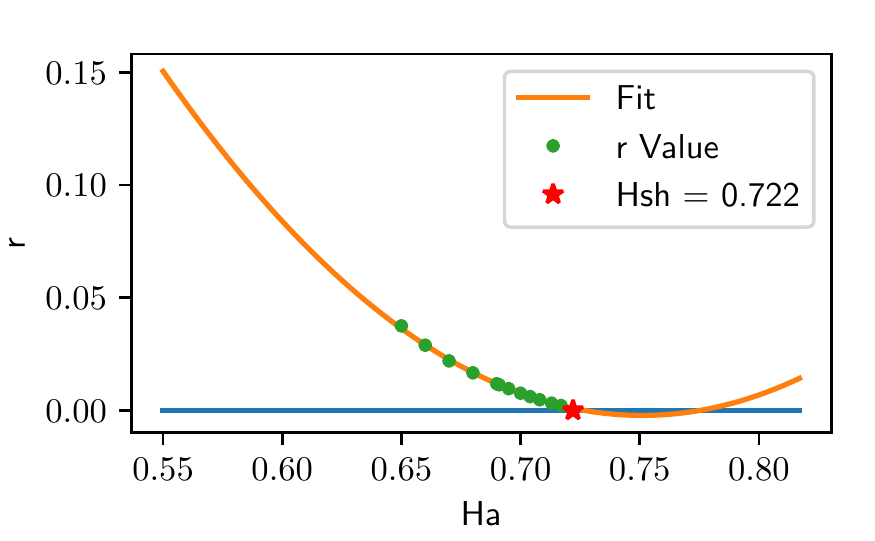}}
    \caption[Ha Symmetric Cylinder]{\label{fig:Symmetric Fit}
      \textbf{Extracting $H_{sh}$.}
      Extrapolating the bifurcation parameter $r$ to zero gives $H_{sh}$.
    }
\end{figure}

The procedure for calculating \Hsh\ and $k_c$ differ from those based on linear-stability analysis in the time-independent case\cite{transtrum2011superheating}.
Here, using bifurcation analysis, we extract the numerical value of the bifurcation parameter $r$ using the observed decay rate of the critical mode.
Repeating this for several different applied fields gives an empirical relationship between $r$ and $H_a$, represented in Figure~\ref{fig:Symmetric Fit}.
The superheating field occurs at $H_a$ such that $r = 0$.
We estimate \Hsh\ by fitting empirical estimates of $r(H_a)$ to a second-order polynomial and solving for $r = 0$.
We also calculate $k_c$ by counting the number of sign changes in the critical mode in Fig. \ref{fig:Symmetric mode}.
Table \ref{table:HshKappa} summarizes our calculations of $H_{sh}$ and $k_c$ for varying $\kappa$ and compares them to previous estimates from\cite{transtrum2011superheating}.

\begin{table}
  \begin{center}
    \begin{tabular}{|l|l|l|l|}
      \hline
      & Cylinder $H_{sh}$ & Slab $H_{sh}$ & Absolute Relative Difference\\ \hline
      $\kappa=2$ & 0.803 & 0.7980 & 0.00615\\ \hline
      $\kappa=4$ & 0.721 & 0.7233 & 0.00320\\ \hline
      $\kappa=6$ & 0.683 & 0.6879 & 0.00711\\ \hline
      $\kappa=8$ & 0.660 & 0.6663 & 0.00944\\ \hline
      & Cylinder $k_c$ & Slab $k_c$ & \\ \hline  
      $\kappa=2$ & 0.975 & 1.1423 & 0.1465\\ \hline
      $\kappa=4$ & 2.125 & 2.31769 & 0.0831\\ \hline
      $\kappa=6$ & 3.125 & 3.27868 & 0.0468\\ \hline
      $\kappa=8$ & 3.925 & 4.15077 & 0.0544\\ 
\hline
    \end{tabular}
  \end{center}
  \caption{ \textbf{Numerical Results.}
    $H_{sh}$ and $k_c$  for different values of $\kappa$ calculated using bifurcation analysis with a cylinder of radius 40.
    For comparison, we include estimates from time-independent calculations.}
\label{table:HshKappa}  
\end{table}

In addition to linear stability analysis, previous work has also used the time-dependent theory to estimate the entry field\cite{soininen1994stability, vodolazov2000effect, burlachkov1991bean, aladyshkin2001best, machida1993direct, koshelev2016optimization, du1994finite, oripov2019time, dorsey1992vortex, li2015new, gao2015efficient, sadovskyy2015stable,sorensen2017dynamics,deang1997vortices,benfenati2019vortex}.
An advantage of using the time-dependent theory, is the ability to explore rough geometries.
Typically, the field is raised until vortices nucleate, but efficiently and accurately determining the transition point can be tedious as the relevant time scales diverge near $H_{sh}$.
The bifurcation analysis we describe above extracts the same information without having to explicitly nucleate vortices.
In the next section we demonstrate qualitative agreement to previous studies.

\subsection{Random Surfaces}

\begin{figure}
    \centerline{\includegraphics[width=1.0\columnwidth]{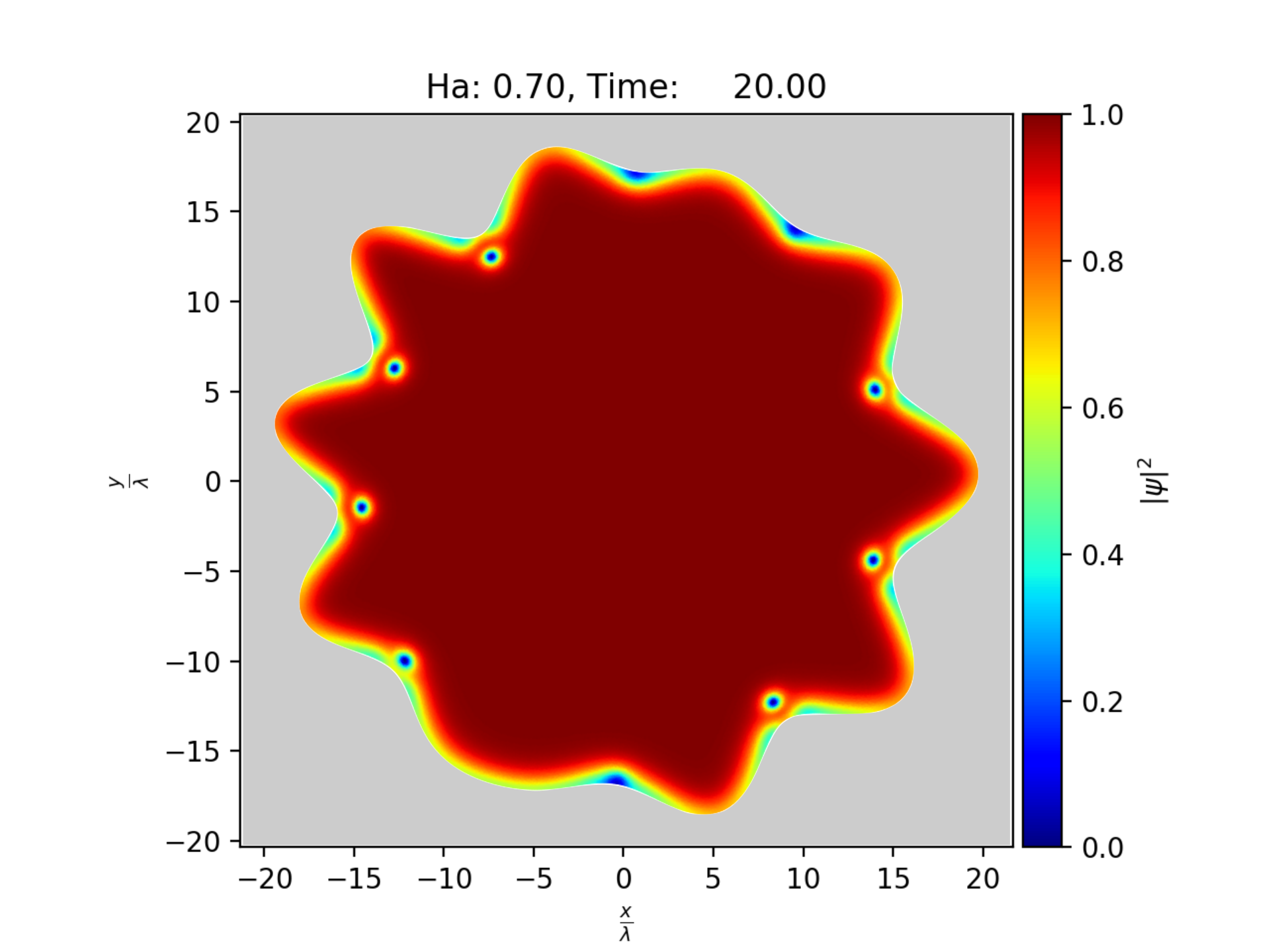}}
    \caption[Rough Pen]{\label{fig: Rough Pen}
      \textbf{Vortex Nucleation for Rough Surfaces.}
      The norm squared of the order parameter just after vortex nucleation.
      Note how the vortices penetrate in the troughs of the surface.
      }
\end{figure}

\begin{figure}
    \centerline{\includegraphics[width=1.0\columnwidth]{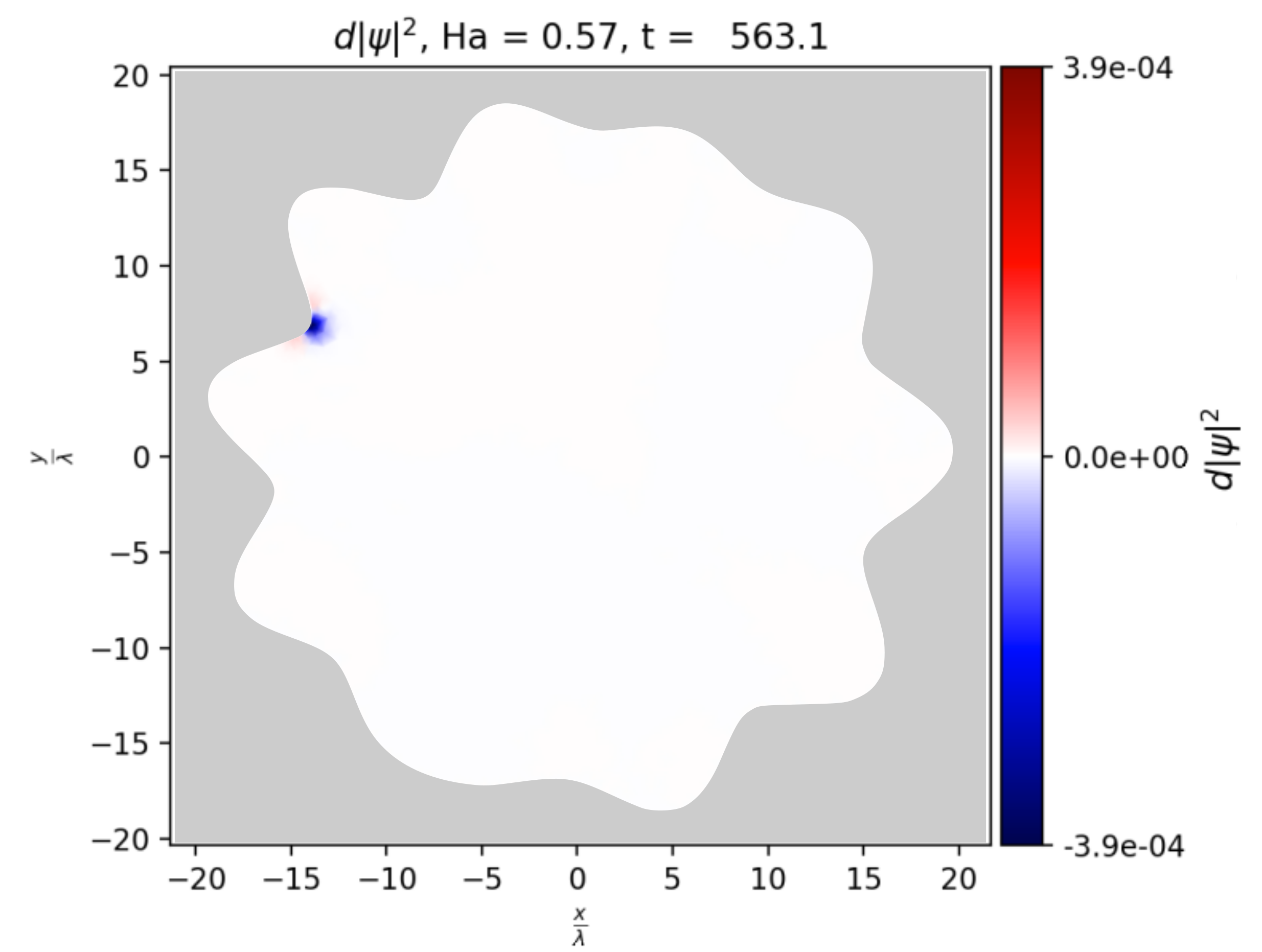}}
    \caption[Rough Pen Mode]{\label{fig: Rough Pen Mode}
      \textbf{Critical Mode for the Rough Surface.}
      The slowest decaying mode for rough surfaces is concentrated at the troughs where the first vortex enters. 
  }
\end{figure}

Vortex nucleation is a surface effect; surface roughness changes how vortices nucleate.
Fig. \ref{fig: Rough Pen} shows a simulation that captures vortex nucleation for a random surface.
Note that $H_a=0.7$ for this simulation and is less than \Hsh\ for the symmetric case.
Also note that the critical fluctuation is no longer a periodic array.
Instead the mode is large at concave regions of the surface, where the vortices first form (See Fig. \ref{fig: Rough Pen Mode}).
Using bifurcation analysis we calculate \Hsh$=0.566$ for this geometry, a significant reduction in in the value for a smooth surface (\Hsh$=0.72$).

\begin{figure}
    \centerline{\includegraphics[width=1.0\columnwidth]{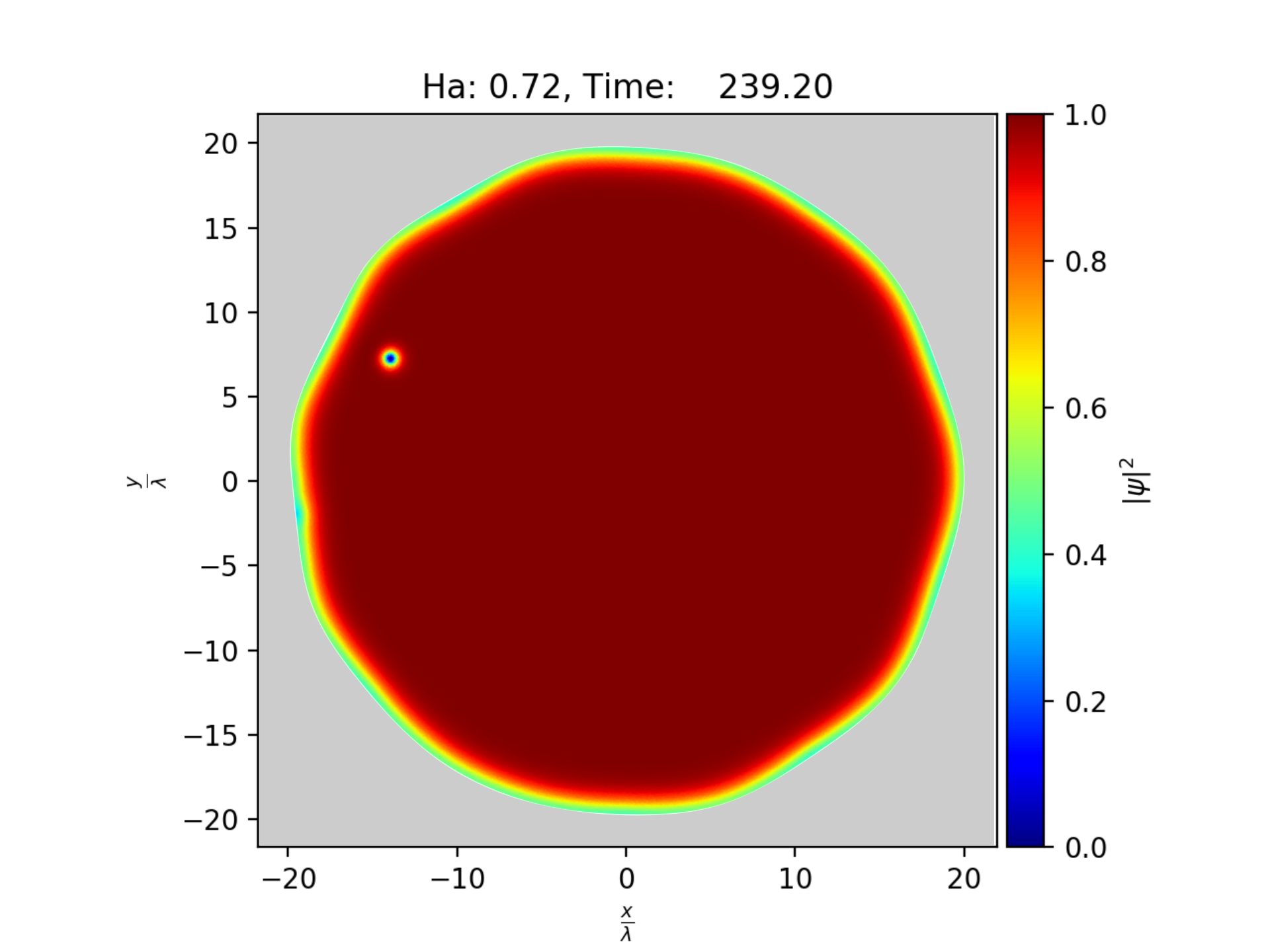}}
    \caption[Small Rough Pen]{\label{fig: Small Rough Pen}
      \textbf{Vortex Nucleation for Small Roughness.}
      Even a little roughness qualitatively changes vortex nucleation pattern.
      Here only one vortex nucleates.
    }
\end{figure}

\begin{figure}
    \centerline{\includegraphics[width=1.0\columnwidth]{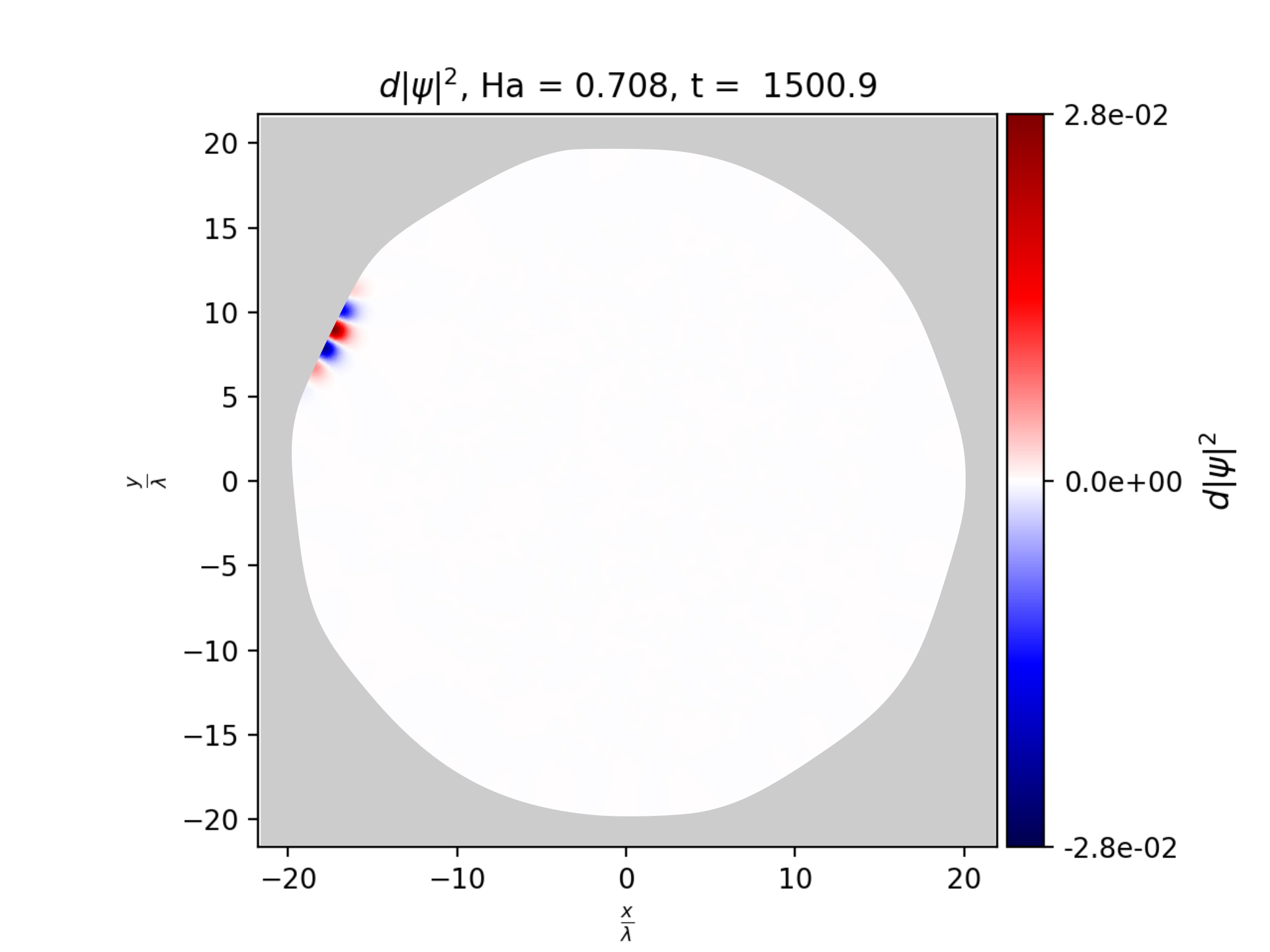}}
    \caption[Small Rough Mode]{\label{fig: Small Rough Mode}
      \textbf{Critical Mode for Small Roughness.}      
      The critical mode is centered where the first vortex enters.
      Compare with Fig.\ref{fig: Small Rough Pen}.
    }
\end{figure}

The roughness in Figure~\ref{fig: Rough Pen} is somewhat extreme, but illustrates the relevant physics in qualitative agreement with previous results.
Although, less roughness leads to smaller reduction in \Hsh, we find that even a very small roughness leads to a large, qualitative change in the critical mode.
Indeed, even very small, individual divots act as nucleation points for vortices, as illustrated in Figures~\ref{fig: Small Rough Pen} and \ref{fig: Small Rough Mode}.

\subsection{Single Divot}

\begin{figure}                  
    \centerline{\includegraphics[width=1.0\columnwidth]{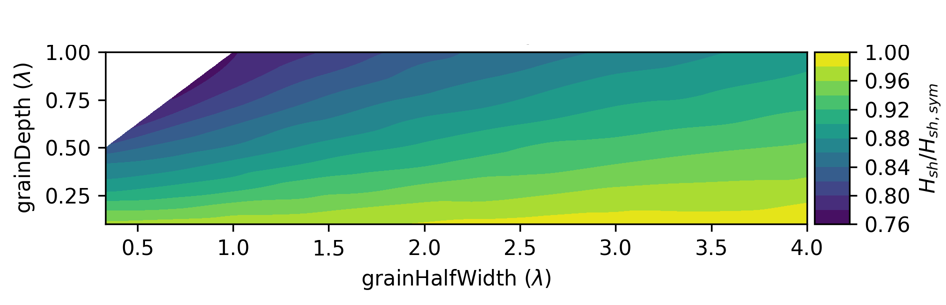}}
    \caption[Grain Heatmap]{\label{fig:Grain Heatmap}
      \textbf{Role of Geometry in Vortex Nucleation.}
      The ratio of \Hsh\ in the presence of a divot to the bulk value.
      Divots that are thin and deep are the most detrimental.
    }
\end{figure}

It has long been known that surface roughness is a relevant parameter for vortex nucleation within GL theory.
To explore which geometric properties affect nucleation, we introduce a single exponential cut out on the surface of the cylinder.
We vary the height and depth of this defect and calculate the corresponding reduction in \Hsh.
Results are summarized in Fig.~\ref{fig:Grain Heatmap}; divots that are narrow and deep lead to the largest reduction in \Hsh.
A similar study assuming large $\kappa$ and using London theory also found single divots to be detrimental\cite{burlachkov1991bean}.

\begin{figure}
    \centerline{\includegraphics[width=1.0\columnwidth]{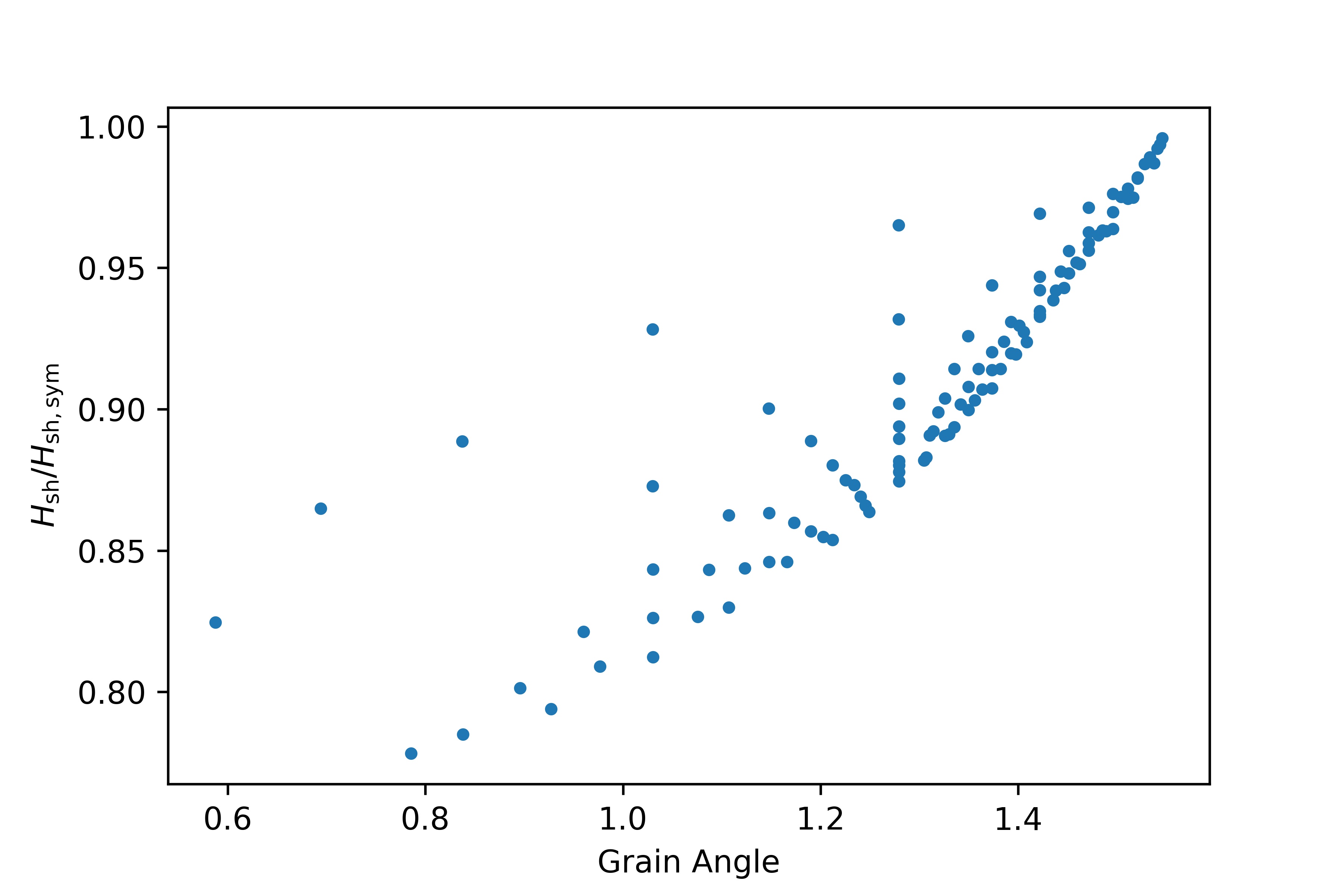}}
    \caption[Grain Heatmap]{\label{fig:Grain Angle}
      \textbf{\Hsh vs. Opening Angle.}
      Divots with the same opening angle may nucleate vortices at different applied fields. 
    }
\end{figure}

An alternative parametrization of the divot geometry is in terms of the opening angle.
A potential hypothesis is that the opening is the relevant parameter determining vortex nucleation; however, Figure~\ref{fig:Grain Angle} shows that this is not the case.

\subsection{Variations of $T_c$}
\begin{figure}
  \centerline{\includegraphics[width=1.0\columnwidth]{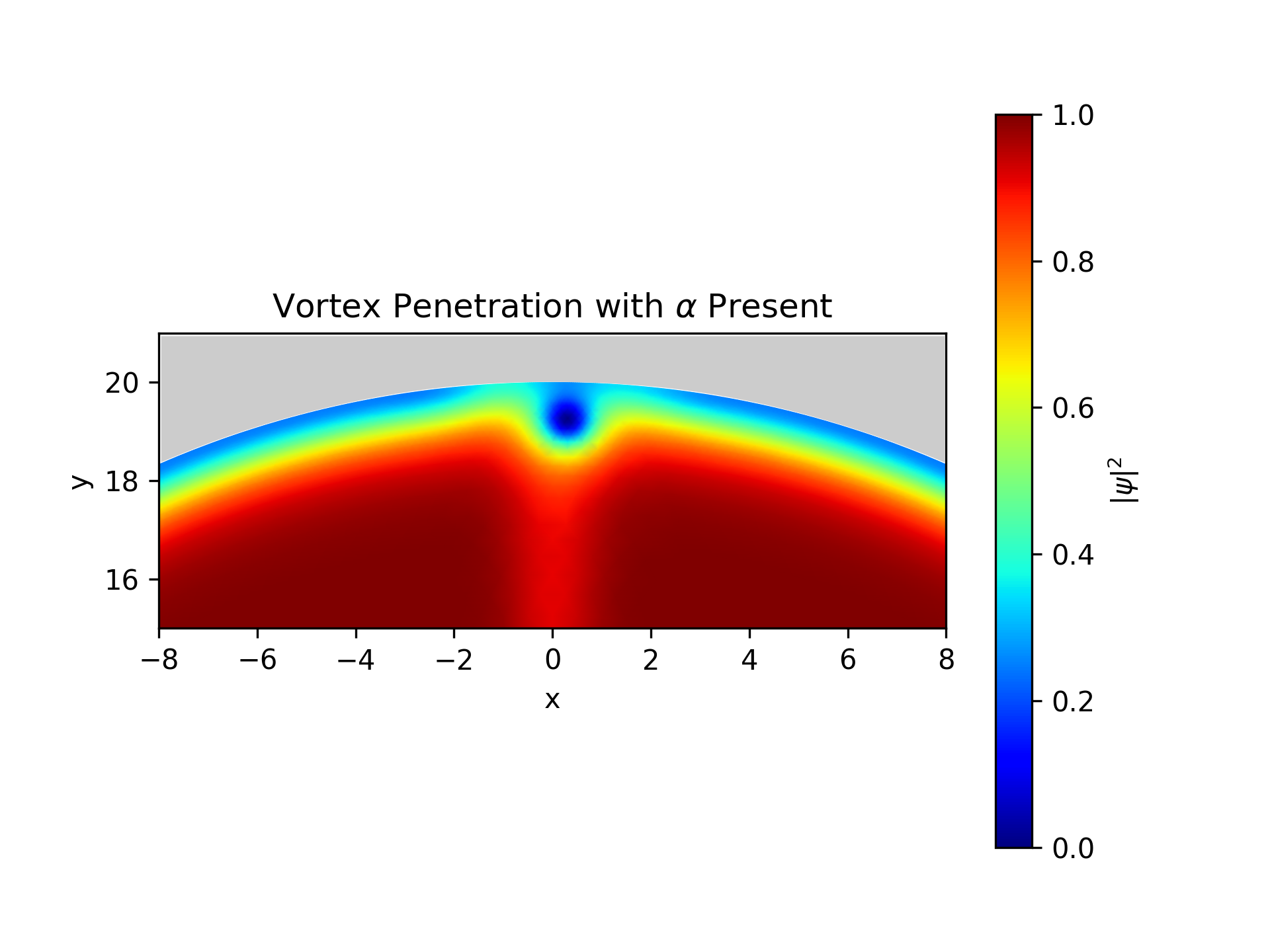}}
  \caption[alpha Nucleation]{\label{fig:alpha Nucleation}
    \textbf{Nucleation Due to Material Inhomogeneity.}
    We plot the norm squared of the order parameter above $H_{sh}$ when $a(r)$ varies as shown in Fig.~\ref{fig:Alpha Cylinder}.
    Vortices nucleate in regions of low $a\propto 1 - T/T_c$ (i.e., lower $T_c$).    
  }
\end{figure}

In addition to surface roughness, material inhomogeneities also act as nucleation sites.
We model variations in material properties by spatially varying $\alpha(r) \propto 1 - T/T_c$ as described in section \ref{Inhomogeneities}.
Fig.~\ref{fig:alpha Nucleation} shows that for $H_a > H_{sh}$ vortices first nucleate where $T_c$ is lowest on the surface.
Similar to surface roughness, even a small, local reduction in $T_c$ leads to a localization in the critical mode.

\begin{figure}
  \centerline{\includegraphics[width=1.0\columnwidth]{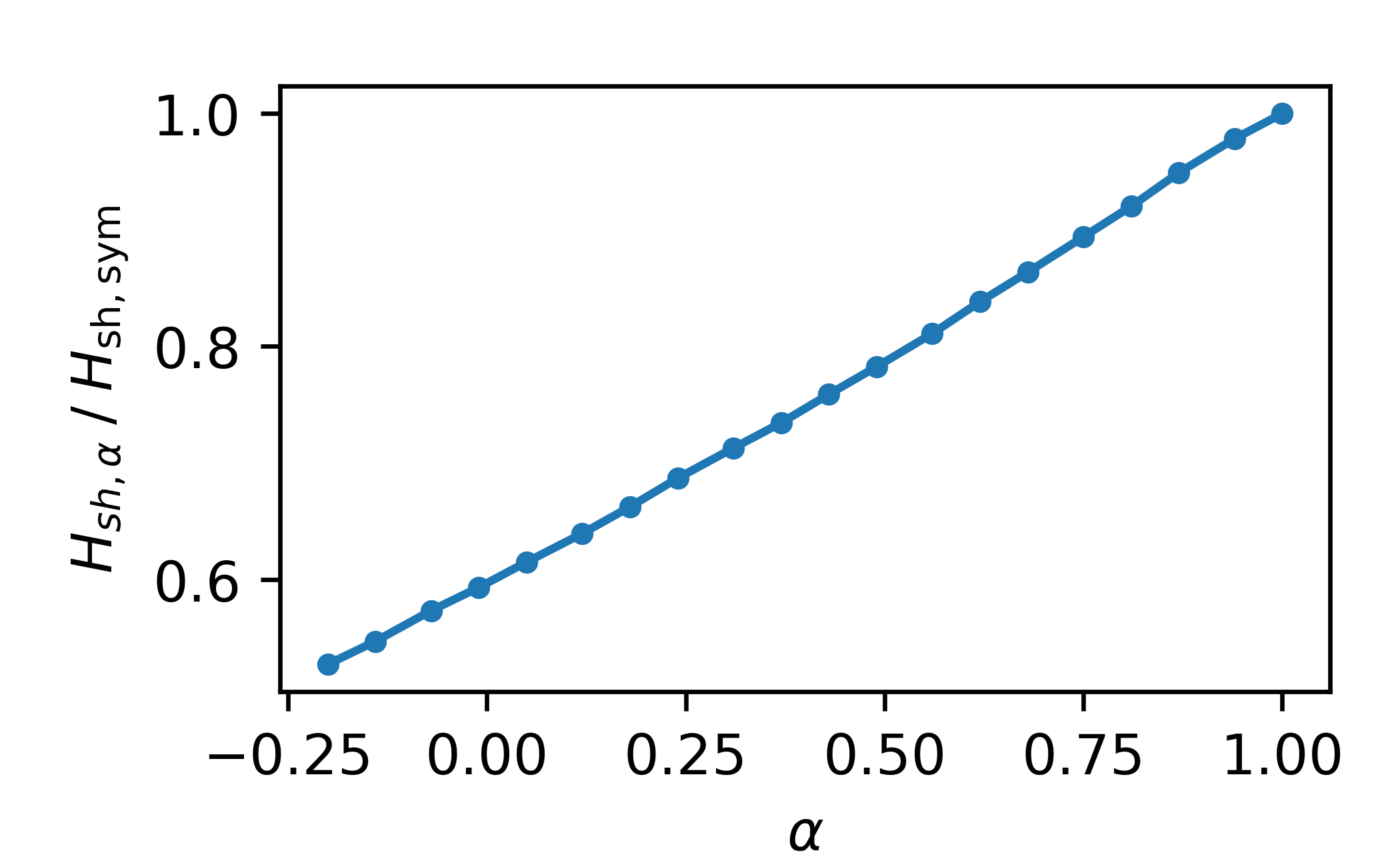}}
  \caption[Hsh vs alpha]{\label{fig:Hsh vs alpha}
    \textbf{Reduction in \Hsh\ vs.~Material Parameter}
    The minimum value of $\alpha$ in the weakly-superconducting region determines the field at which vortices first nucleate.
  }
\end{figure}

Variation in $T_c$ can also lead to a significant drop in $H_{sh}$ as seen in Fig.~\ref{fig:Hsh vs alpha}. 
As a point of comparison, for Nb$_3$Sn, the variation in Sn concentration can cause $T_c$ to vary from about 18K at the optimal stoichiometry to as low at 6K in Sn depleted regions Sn seen in typical SRF cavities\cite{devantay1981physical,devantay1982superconductivity,godeke2006review,becker2015analysis, lee2018atomic, sitaraman2019textit}.
For an SRF cavity operating near 4K, this means that vortices could nucleate at an applied field around 0.6, an effect comparable to the extreme roughness of Figure~\ref{fig: Rough Pen}.
These results suggest that realistic variations in $T_c$ could be an important mechanism for vortex nucleation.

\subsection{Film Geometry}

Up to this point, all our results have been reported for the two-dimensional cylindrical geometry.
To control for the effects of curvature, we repeat our calculations using a film geometry.
We apply the same magnetic fields to the top and bottom of the rectangular domain and enforce periodic boundary conditions on the left and right sides.
Our results for the film geometry are nearly identical to those of the cylinder, indicating that the curvature effects are minimal.

\subsection{3D Film}

A major limitation of the two-dimensional analysis is that the magnetic field must be parallel to the defects.

As mentioned in section \ref{Problem Formulation}, the 2D geometry is a cross section of a 3D domain that does not vary in the $\hat{z}$ direction.
In this geometry it is not possible to simulate defects that break symmetries in the direction the magnetic field points, nor it is possible to have defects oriented differently from the applied field.
To consider magnetic fields perpendicular to defects, we must move into fully three-dimensional geometries.
Because three-dimensional simulations are more computationally expensive, we only consider volumes that accommodate a single vortex.
Our geometry is a three-dimensional generalization of the 2D film.
We fix the applied field on the faces parallel to the z-plane and apply periodic boundary conditions to the remaining sides.
We use a mesh that is 2$\lambda$ long in the x direction, 1.5$\lambda$ in the y, and 5$\lambda$ in the z direction.

\begin{figure}
  \centerline{\includegraphics[width=1.0\columnwidth]{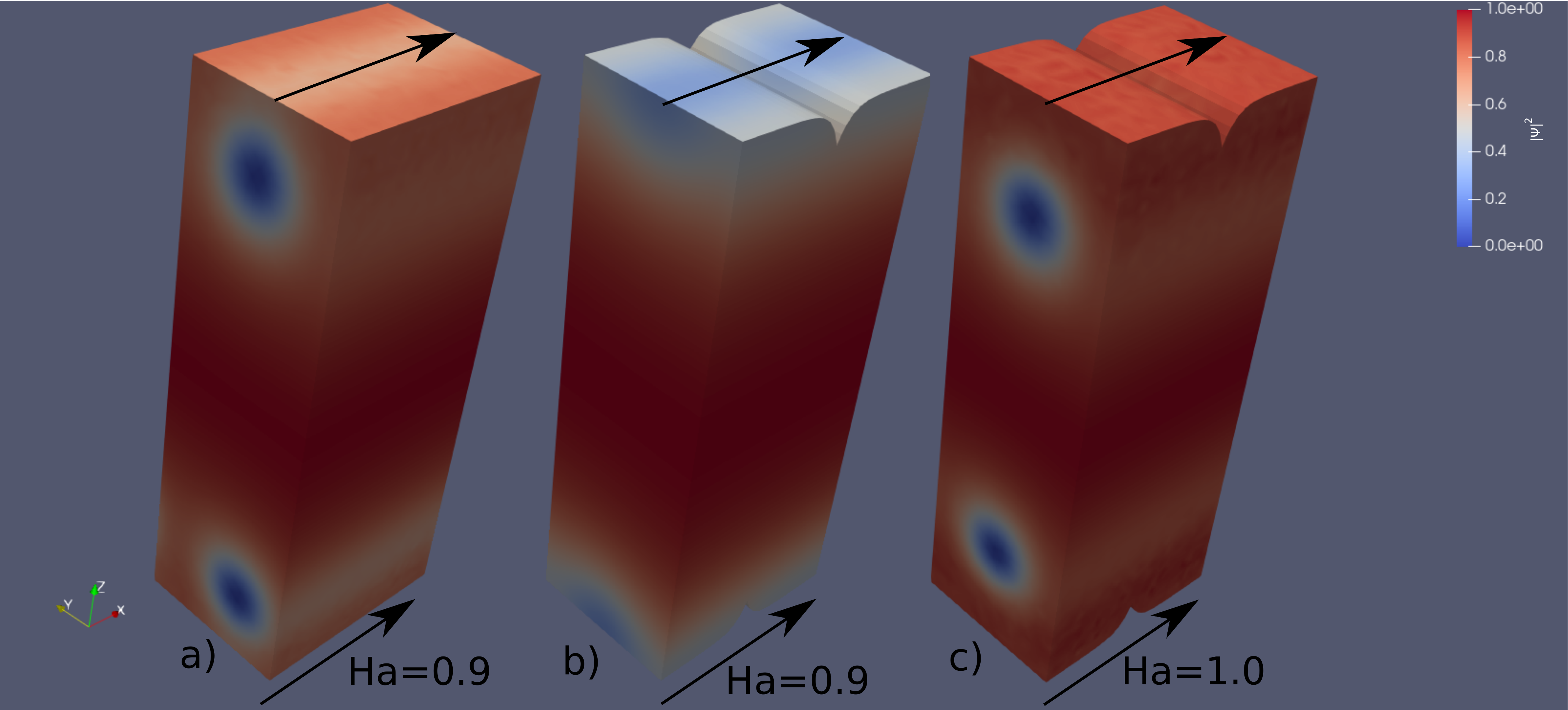}}
  \caption[3D all]{\label{fig:3D all}
    \textbf{Vortex Nucleation in 3D.}
    Plotting the square of the order parameter shows that a vortex has nucleated at $H_a = 0.9$ for the smooth surface but not the dented surface.
    At $H_a=1.0$ we do see vortex nucleation perpendicular to the divot. 
  }
\end{figure}

Our results indicate that when defects are perpendicular to the applied field the superheating field is effectively raised.
We illustrate in Fig.~\ref{fig:3D all} in which we observe a vortex nucleating on a smooth surface at an applied field of $H_a = 0.9$.
The magnetic field direction is indicated by the black arrow.
However, after introducing a defect perpendicular to the magnetic field, no vortex nucleates at $H_a = 0.9$.
After raising the field to $H_a = 1.0$, the vortex fully enters the dented film.
This demonstrates that the relative orientation of defects and the applied also plays a crucial role in nucleation mechanism and suggests that the most dangerous divots are those parallel to the applied magnetic field.

\section{Discussion and Conclusions}
\label{sec:Summary}

This work combines TDGL simulations with bifurcation analysis to study the transition of the metastable Meissner state to the mixed state of type II superconductors.
We have implemented a finite-element method that accommodates rough geometries in two- and three-dimensions, as well as variations in material parameters.
We have demonstrated accuracy by reproducing previous calculations of $H_{sh}$ and $k_c$ for smooth geometries.
The flexibility of finite element methods enable simulating geometries that are more complex, including both rough surfaces and material inhomogeneities.
The bifurcation analysis allows us to efficiently extract both the superheating field as well as the accompnying critical mode without explicitly simulating vortex nucleation which occurs at diverging time scales.

We have shown that even very small surface roughness and material inhomogeneity can change the nucleation mechanism.
In smooth geometries, arrays of vortices nucleate together.
However, weak perturbations lead to a localization of the critical model and significant reduction in \Hsh.
Future work will further apply these tools to geometries and material-specific parameters motivated by experimental observations.

As we are interested in defects about the size of a coherence length we focus on mesoscopic scales.
We've chosen the penetration depth as our length scale in our simulations.
This means that for increasing $\kappa$ we must consider smaller coherence lengths.
This further increases mesh density and makes simulations more computationally expensive.
The value of $\kappa$ determines how large of a domain we can simulate.
For type-II materials, such as Nb$_3$Sn, simulations will be primarily limitd to mesoscopic scales.

This work has been based on Ginzburg-Landau theory that has known limitations.
Most importantly, GL theory is formally exact only when the system is close to its critical temperature; however, most SRF cavities operate well below $T_c$.
Previous work applying Eilenberger theory to uniform surfaces and materials suggests that the Ginzburg-Landau predictions are surprisingly accurate (within a few percent) even at very low temperatures\cite{catelani2008temperature}.
It is reasonable to expect that the \emph{relative} effects of roughness and material inhomogeneity that we have quantified will hold even at low temperatures, and that inhomogeneities are likely to be bottlenecks to performance. 

A critical aspect that we have ignored here is field enhancement.
The field enhancement effect refers to a local increase in the applied field in response to surface roughness.
Our simulations have not accounted for any field enhancement effects.
This would require solving Maxwell's equations in the vacuum region outside the superconductor.
This could be added in future work, but is beyond the scope of this study.

This analysis is a step toward sample-specific time-independent calculations of \Hsh\ that includes not only surface defects, but spatially varying material parameters.
We have shown that realistic variations in $T_c$ can lower the barrier to vortex nucleation in ways similar to surface roughness and such effects are likely to be present in alloyed superconductors.
We present these results as an exploration of GL theory and as a tool for quantifying detrimental defects in realistic superconducting samples.
In the future we plan to extend these results to incorporate more material parameters and specific geometries into this framework and how these tools are bringing insight to the development of Nb$_3$Sn cavities.

We thank James Sethna, Danilo Liarte, Matthias Liepe, Tomas Arias, Sam Posen, Richard Hennig, Nathan Sitaraman, Michelle Kelley, Aiden Harbick, and Braedon Jones for helpful discussions.
This work was supported by the US National Science Foundation under Award OIA-1549132, the Center for Bright Beams.

\appendix*
\section{Meshing}

We simulate 3 geometries, the 2D cylinder, the 2D film, and the 3D film.
In all of these geometeries the mesh is refined to capture length scales smaller than the order parameter, otherwise the simulations do not accurately capture vortex dynamics.

\begin{figure}
    \centerline{\includegraphics[width=0.9\columnwidth]{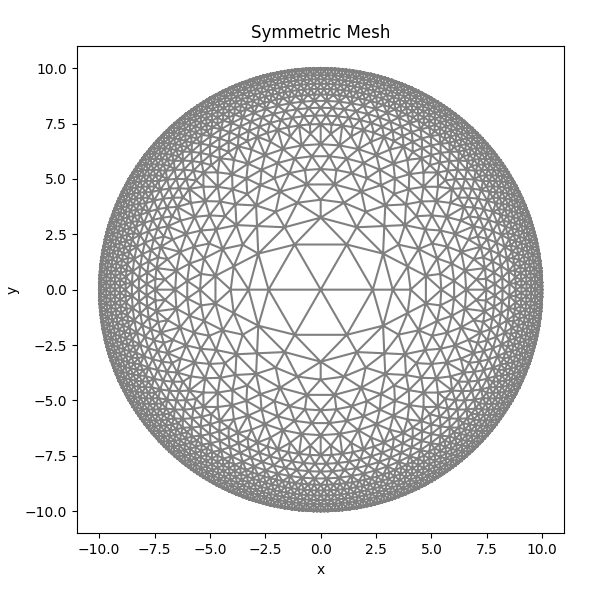}}
    \caption[Symmetric Cylinder Mesh]{\label{fig:Symmetric Cylinder}
      We solve the time-dependent Ginzburg-Landau equations on a circular cutout of a cylinder.
      Forcing symmetry in the mesh insures vortices penetrate uniformly.
      We refine the mesh near the surface as we are only interested in initial vortex nucleation.
      Length is measured in penetration depths}
\end{figure}

For the smooth cylinder we want to keep the simulation as symmetric as possible to minimize the effect of numerical noise.
Near \Hsh\ small defects in the mesh can lead to vortex nucleation.
For this reason we divide the domain into concentric circles.
Starting with the inner circle we add points equally around the circumference.
We then add points to the second largest circle such that if projected onto the inner circle they would be centered between the first set of points.
We repeat this process adding extra points if the domain becomes too sparse.
Finally, we are interested in dynamics near the surface so we push interior points radially outward.
Fig.\ref{fig:Symmetric Cylinder} shows the end result of this process for a cylinder of radius 10$\lambda$.

\begin{figure}
    \centerline{\includegraphics[width=0.9\columnwidth,trim=2.0cm 0.0cm 3.0cm 0.75cm,clip]{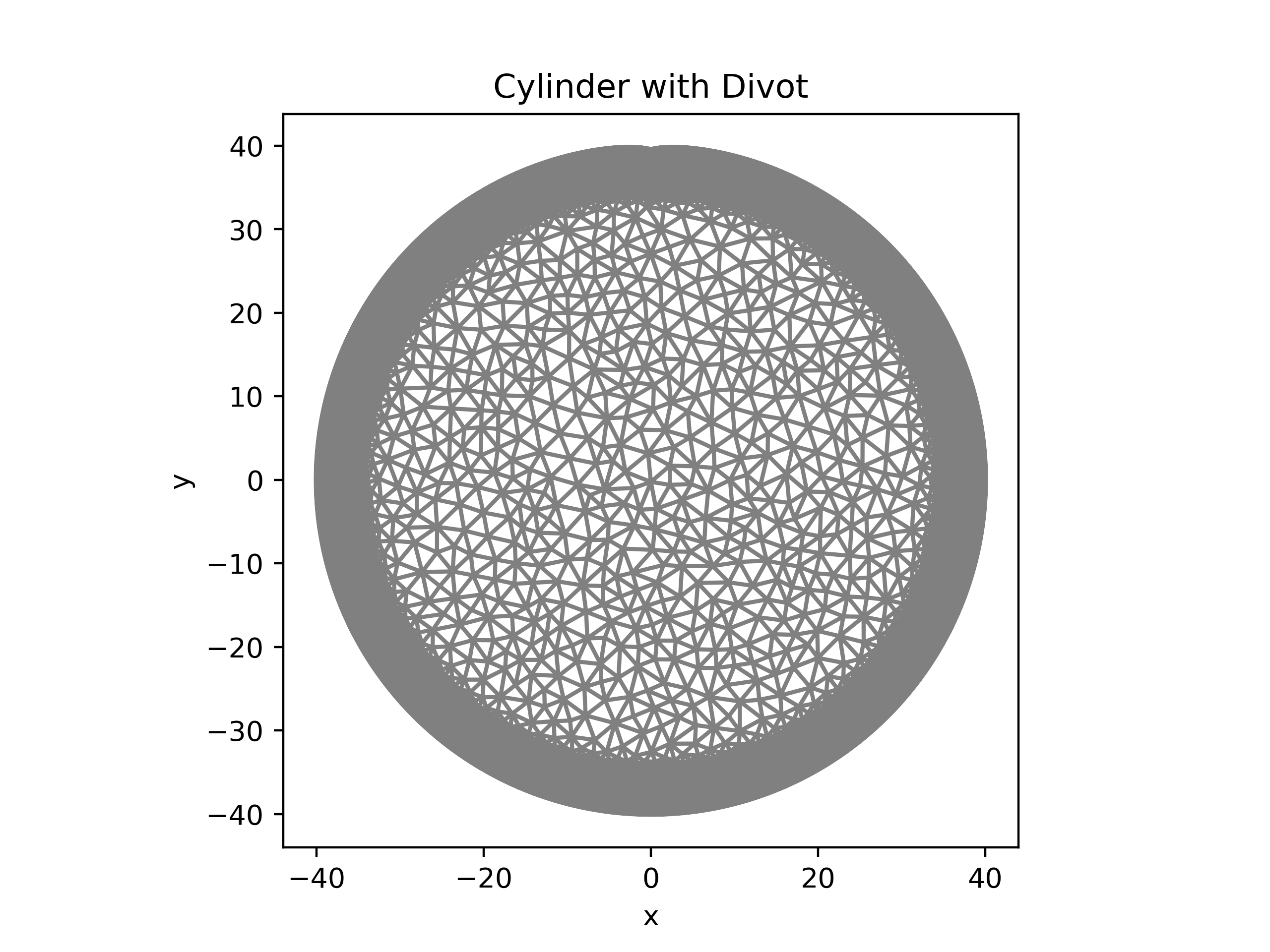}}
    \caption[Grain Boundary Mesh]{\label{fig:Grain Boundary Geometry}
      We introduce a geometric defect to the cylinder based on experimentally observed grain boundaries.}
\end{figure}

\begin{figure}
    \centerline{\includegraphics[width=0.9\columnwidth,trim=0.25cm 3.25cm 1.0cm 4.0cm,clip]{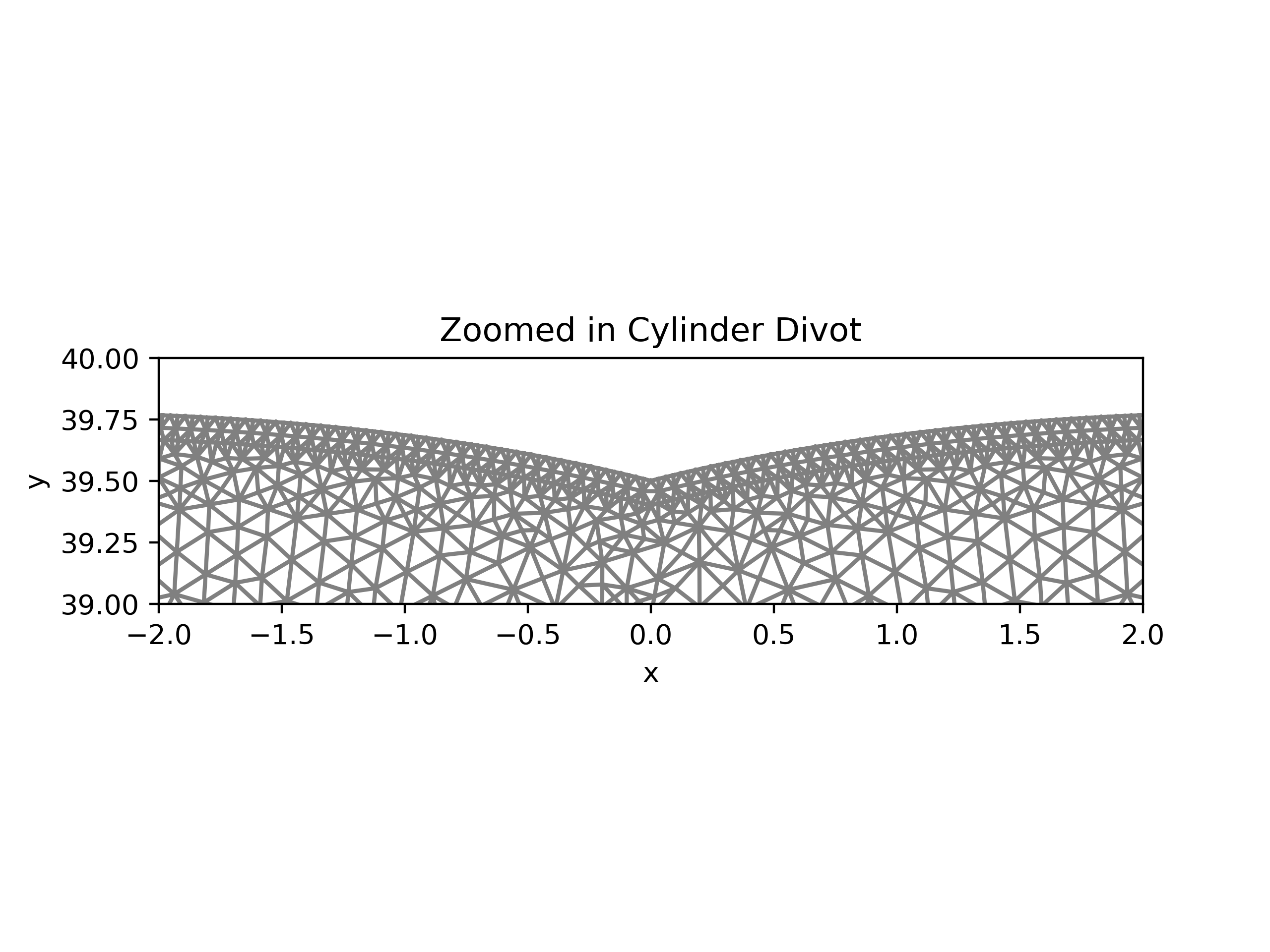}}
    \caption[Grain Boundary Close]{\label{fig:Grain Boundary Close}
      We etch out an exponential-like function from the surface of our cylinder to match what is observed experimentally.}
\end{figure}

Once we introduce an inhomogeneity the local defect dominates global behavior.
It is no longer necessary to keep the mesh symmetric.
We can let FEniCS automatically mesh the domain.
We can define differing mesh densities for different regions as in Fig.\ref{fig:Grain Boundary Geometry}.
In Fig.\ref{fig:Grain Boundary Close} we can see the mesh close to the defect.

\begin{figure}
    \centerline{\includegraphics[width=0.9\columnwidth,trim=2.25cm 0.25cm 3.0cm 0.75cm,clip]{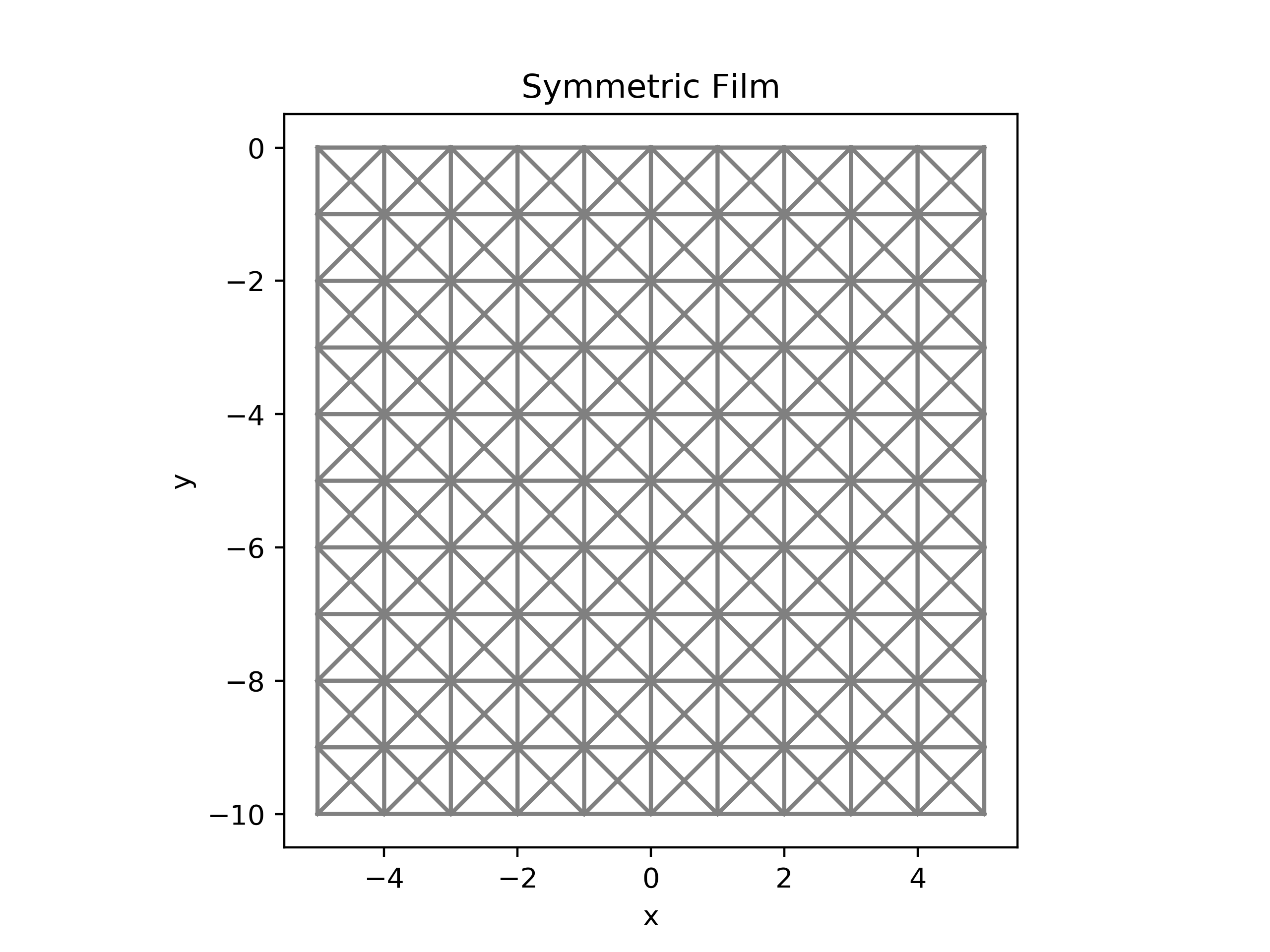}}
    \caption[Grain Boundary Close]{\label{fig: Sym Film}
      The symmetric mesh is broken up into smaller rectangles containing 4 triangles.
      This prevents biases in vortex movement.
    }
\end{figure}

\begin{figure}
    \centerline{\includegraphics[width=0.9\columnwidth,trim=2.25cm 0.25cm 3.0cm 0.75cm,clip]{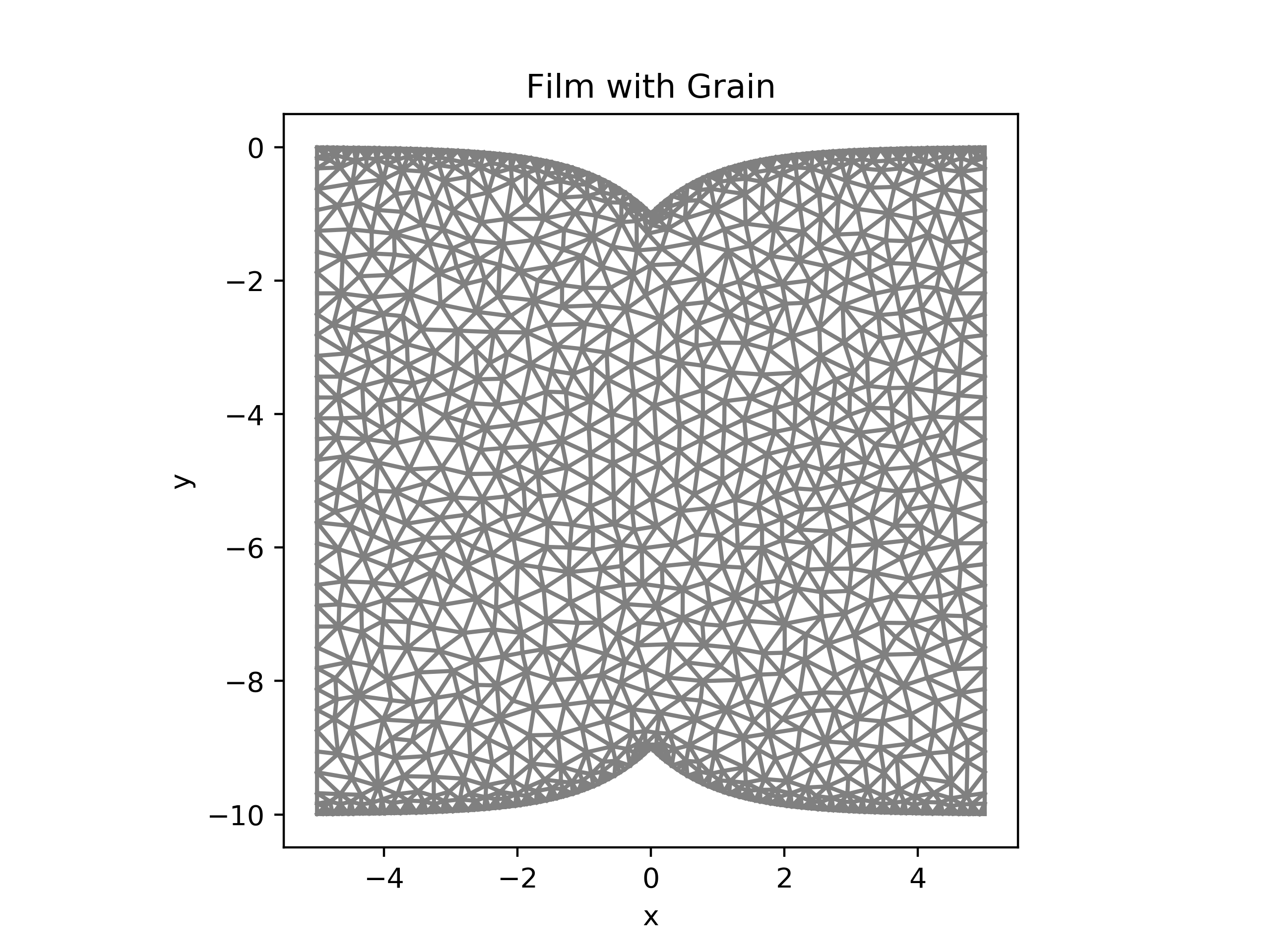}}
    \caption[Grain Boundary Close]{\label{fig: Grain Film}
      Here we add a divot to the surface of the film mesh.}
\end{figure}

As a reference for future papers here is how we mesh the film.
The domain is broken up into rectangles.
We found that if we split the rectangles into an upper right triangle and a lower left triangle then nucleated vortices came in at an angle.
To avoid this we divide each rectangle into 4 triangles as seen in the Fig. \ref{fig: Sym Film}.
When we introduce a divot the surface gets remeshed and this bias disappears as seen in Fig. \ref{fig: Grain Film}

In 3D we only considered a domain that was big enough for 1 vortex to form.
The surface has a symmetric grid of points.
When we introduced a defect we centered the cusp on a line of vertex points. 
Interior points were not symmetric.

\bibliography{TDGLPaperrefs}

\end{document}